# The 2018 Martian Global Dust Storm over the South Polar Region studied with MEx/VMC

J. Hernández-Bernal[1,2], A. Sánchez-Lavega[1], T. del Río-Gaztelurrutia[1], R. Hueso[1], A. Cardesín-Moinelo[3,4], E. Ravanis[3], A. de Burgos-Sierra[3], D. Titov[4], S. Wood[5]

[1]Dpto. Física Aplicada I, EIB, Universidad País Vasco UPV/EHU, Bilbao, Spain [2]Aula EspaZio Gela, Escuela de Ingeniería de Bilbao, Universidad del País Vasco UPV/EHU, Bilbao, Spain [3]European Space Agency, ESAC, Madrid, Spain [4]Instituto de Astrofísica e Ciências do Espaço, Obs. Astronomico de Lisboa, Portugal [4]European Space Agency, ESTEC, Noordwijk, The Netherlands [5]European Space Agency, ESOC, Darmstadt, Germany

**Corresponding author:** Jorge Hernández-Bernal (jorge.hernandez@ehu.eus)

**Key Points:**

- The 2018 Global Dust Storm propagated unevenly over the South Polar Region, not covering it fully, and forming elongated narrow dust arcs.
- Overall, dust moved towards the terminator, reaching velocities up to 100 ms-1 in the morning side.
- During June-July 2018, the top altitude of dust showed both spatial and temporal variability, ranging from 10 – 70 km.

## Abstract

We study the 2018 Martian Global Dust Storm (GDS 2018) over the Southern Polar Region using images obtained by the Visual Monitoring Camera (VMC) on board Mars Express during June and July 2018. Dust penetrated into the polar cap region but never covered the cap completely, and its spatial distribution was nonhomogeneous and rapidly changing. However, we detected long but narrow aerosol curved arcs with a length of ~2,000 – 3,000 km traversing part of the cap and crossing the terminator into the night side. Tracking discrete dust clouds allowed measurements of their motions that were towards the terminator with velocities up to 100 ms-1. The images of the dust projected into the Martian limb show maximum altitudes of ~ 70 km but with large spatial and temporal variations. We discuss these results in the context of the predictions of a numerical model for dust storm scenario.

## Plain Language Summary

Dust storms of different scales (local, regional...) are common on Mars. Some Martian years a regional storm activates secondary storms and dust encircles the planet, in a dust event usually called a Global Dust Storm. The last Global Dust Storm took place in 2018, and we are not currently able to predict when a new one will occur. Global Dust Storms affect the global dynamics of the Martian atmosphere, and the dynamics of the Polar Regions is a good proxy to the global situation. In this paper, we take advantage of the polar orbit of Mars Express to study the Southern Polar Region during 2018 Global Dust Storm using the Visual Monitoring Camera onboard the spacecraft. We show how the dust penetrated into the Polar Cap, the apparition of aerosol arcs curved around the pole, and the presence of winds blowing up to 100m/s, not following the usual patterns expected with no Global Dust Storm.

## 1. Introduction

Mars Global Dust Storms (GDS) are uncommon and nowadays unpredictable aperiodic events (Khare et al., 2017, Montabone and Forget, 2018, and references therein for recent reviews). On 30 May 2018 (Martian Year MY 34), a Dust Storm started in Acidalia Planitia, rapidly evolving to become a planet encircling dust storm (GDS 2018), as described from ground-based observations (Sánchez-Lavega et al., 2019) and from in situ measurements by Curiosity rover (MSL) (Guzewich et al., 2019). Ground-based images show that the storm reached the South Polar Cap edge at 60ºS on 6 June, and that by mid-June dust had penetrated the Polar Cap (Sánchez-Lavega et al. 2019). However, the nearly equatorial viewing geometry from ground-based telescopes prevented a detailed study of the dust propagation in the Southern Polar Region (SPR) and its interaction with the South Polar Cap.

Mars Express (MEx) polar orbit allows a nearly nadir view of the poles, and images of both polar regions are regularly obtained using the Visual Monitoring Camera (VMC) (Ormston et al. 2011; Sánchez-Lavega et al., 2018a). Due to a technical pause in VMC operations, monitoring of the 2018 GDS started on 18 June, approximately 20 sols after its onset (Sánchez-Lavega et al., 2019). During the rest of June, and in July and August, VMC obtained a large set of images showing the SPR (Figure 1) at a typical resolution in the range of 9-13 km/pixel (details on the list of observations are given in the Supporting Information). The period under study spans from 18 June to 3 August 2018 ($L_s$ ~ 195°-220°). The propagation of the GDS over the SPR is a unique opportunity to study the interaction of a planetary-scale phenomenon with the particular dynamics of the polar atmosphere such as the polar vortex, katabatic winds and condensate clouds (Smith et al., 2018). The behavior of dust from GDS 2018 over the SPR is the main subject of study in this paper. In addition, VMC also performed dedicated limb observations when MEx was near pericenter which allowed us to measure aerosol top altitude and horizontal distribution (Sánchez-Lavega et al. 2018a).

For geometry calculations and navigation of images, two software packages were used: PLIA (Hueso et al. 2010), and a new package named Elkano (see Supporting Information). Geometry parameters were extracted using SPICE (Acton 1996; Acton et al., 2018), from publicly available SPICE Kernels provided by the ESA SPICE Service (Costa 2018). MOLA topographic data (Smith et al.

1999) were taken into account when necessary. For time calculations, we use algorithms and definitions described by Allison (1997) and Allison & McEwen (2000). See Supporting Information for further methodological details.

## 2. Dust Distribution in the Southern Polar Region (SPR)

In mid-June 2018, the edge of the Southern Polar Cap (the area of the South Polar Region covered by ice) was at latitude 60°S, in good agreement with previous measurements in this season (Ls ~ 195°) of the MY (Schmidt et al., 2009). Ground based observations indicate that the propagation of the GDS 2018 into the Southern Polar Cap region started on 8-9 June, reaching latitude 70°S at longitudes 0° to 60°E (Sanchez-Lavega et al., 2019). The dust further penetrated on 11 June to latitude 75°S, at two or more different longitudes (0° to 60°E and ~ 225°E), according to a NASA image animation PIA22519, based on image sequences obtained with the Mars Color Imager (MARCI) on board Mars Reconnaissance Orbiter (MRO) (https://photojournal.jpl.nasa.gov/catalog/PIA22519).

VMC started taking images of the SPR on 18 June. The dust distribution was clearly non-homogenous and showed daily changes throughout June (Figure 1b-e). Images taken on the 18 and 23 June show longitude sectors from 270° to 0°E (Figure 1a-b) and 150° to 200° (Figure 1d) respectively. In those images, surface features of the SPR are clearly seen, indicating a limited presence of dust. On 21 and 26 June the presence of thick dust is noticeable at longitudes eastward 90°E (Figure 1c,e). Much clearer regions are seen westwards of Hellas where surface details, such as crater Barnard are easily identified. This region is imaged again on 11 July (Figure 1f) showing that the dust is still present although in smaller quantities. Various surface features can be identified in Figure 1 (panels c,e,f), especially in the 90°-180° range of longitudes, with different degrees of contrast that indicate a variable coverage by dust. Nevertheless, dust rarely hides surface features completely in the southernmost regions. It is worth noticing that morning hazes are usually present about 3-6 hours after sunrise (blue arrows in Figure 1) and mix with the dust. Since they appear in the morning hours to disappear later, these hazes are most likely water ice condensates of relatively low optical depth. In summary, VMC image sequences in June-July show that dust entered the polar cap region, but penetration was irregular, and it never covered the entire polar cap. Additional images of the SPR are presented in Supporting Information Figure S1.

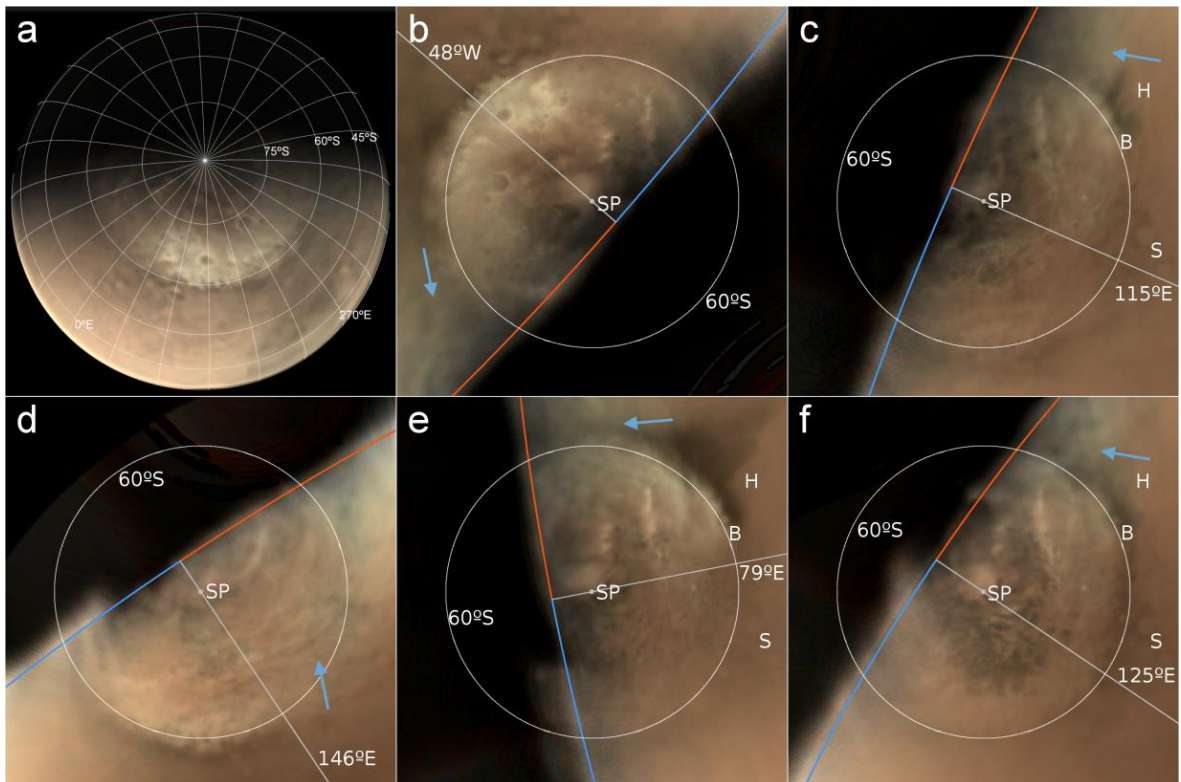

**Figure 1.** Images of the SPR during GDS 2018. (a) Direct polar view: 18 June, 21.50 U.T and its polar projection (b). (b-f) Polar projections covering latitudes 50ºS-90ºS. (c) June 21, 12.40 U.T.; (d) 23 June, 11.50 U.T.; (e) 26 June, 18.25 U.T.; and (f) 11 July, 00.35 U.T. Orange and blue curves indicate respectively the morning and the evening terminator, and the grey line represents the subsolar meridian. In all projections, longitude 0ºE points upwards and east longitude increases clockwise. Hellas planitia, and Barnard and Secchi craters are identified by the letters H, B and S respectively. SP stands for South Pole. The blue arrows indicate morning hazes. See Supporting Information for further details on observations and image processing.

## 3. Arc-Shaped Aerosol Bands over the Southern Polar Region (SPR) seen in Twilight

The season under study follows the southern spring equinox ($L_s$ ~ 195°-220°), hence at any given time, a part of the SPR was not illuminated by the Sun. High altitude aerosols can get direct illumination by the setting Sun in the night, and thus become visible against a dark background. MEx/VMC observations included some high exposure images, which enabled visualizing bright structures in the night side beyond the terminator (Figure 2). These features are organized in one or more arc-shaped parallel bands (Figure 2a), curving around the pole, and extending from the evening terminator to the morning terminator. In some cases the features are not completely visible, with bands captured in the night side of the terminator both in the evening and morning terminators, but not in regions at local midnight (Figure 2a-b). In other cases, the bands are seen in full length (Figure 2c-d). In the morning terminator, the bands are harder to distinguish since they mix with the morning hazes (Figure 2a-b). In Figure 2e we show the location of all measured bands against a map of the SPR.

At the evening terminator, the bands are closer to the pole, at distances ranging between 300 and 1100 km from it. They extend across a typical length of 2000-3000 km, ending further from the pole, at about 500-1500 km in the morning terminator. The bands show irregular morphology with widths in the range ~ 200-500 km. Additionally, an image from 1 July shows two or three parallel curved bands in the dayside, separated by ~ 100-200 km and extending up to a thousand kilometers in length at a mean latitude of 70ºS, 1200 km from the pole (Figure 2a). The visibility of the bands in the night side is an indication of their altitude, which we determine by simple geometry (Hernández-Bernal et al. 2018) to be in the range of 10-35 km ± 5 km, higher in the evening side than in the morning side. The presence of banding becomes more frequent as the season progresses, and after 18 July, at least one band is fully visible in the night in most cases. As can be seen in Figure 2f, this is mainly a consequence of the change in the position of the terminator with the advancing season. Moreover, as the season advances, the bands become less curved, their centers recede from the terminator and they move towards the morning side. Unfortunately, the lack of reliable spectral information from VMC prevents us from determining the nature of the aerosol forming the bands, although it is very likely that this is dust from the GDS. The shape and distribution of the bands suggests that they are organized by the polar atmospheric dynamics. One possibility is that the polar vortex causes the organization of these bands, a possibility discussed in the Conclusions (section 7).

**The 2018 Martian Global Dust Storm over the South Polar Region studied with MEx/VMC**
Hernández-Bernal et al. 2019. Manuscript accepted for publication on Geophysical Research Letters
This document is distributed under CC BY-SA 3.0 IGO license

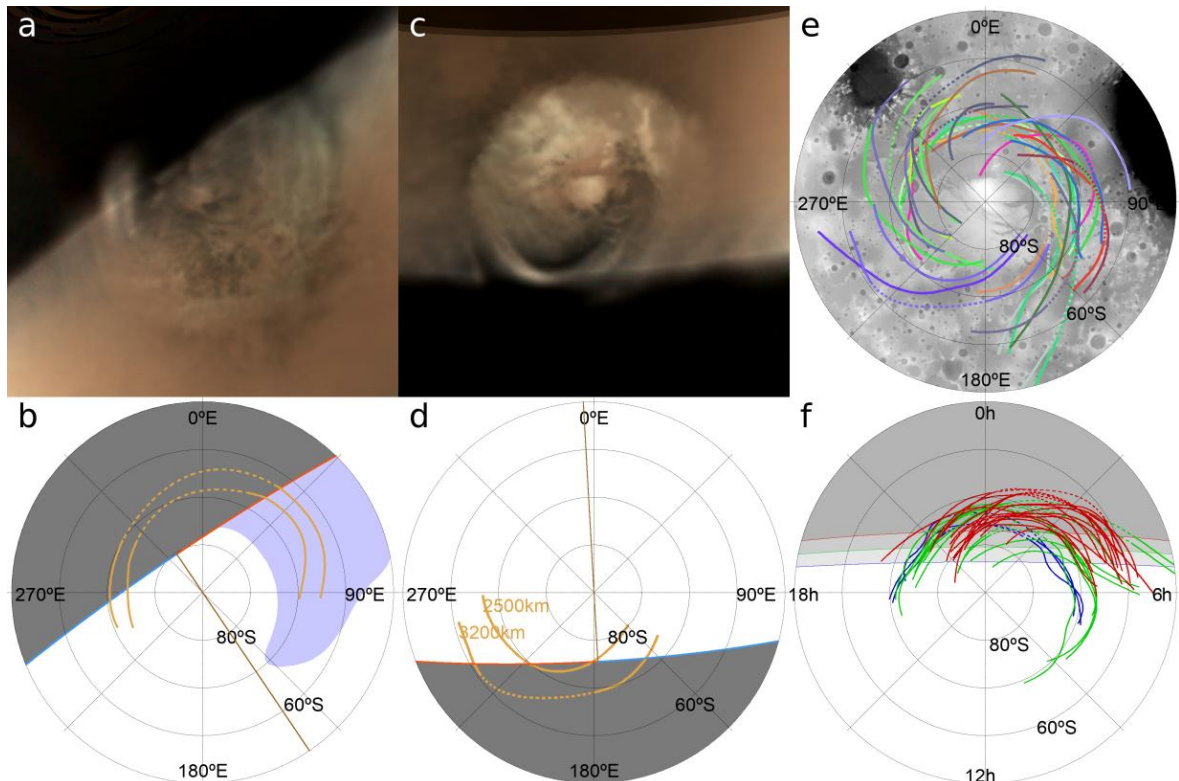

**Figure 2.** Images and structure of the arc bands around the South Pole. (a), (c) Polar projected images of the southern polar region on 1 July, 17:30 U.T. and 22 July, 16:40 U.T. respectively. (b),(d) Schematic representations of previous images showing the night side (dark grey), the morning hazes (blue lilac), and the observed bands, with continuous orange lines indicating visible parts and dotted sections indicating the potential location in the night side, Numbers in orange indicate the estimated length of arcs. Red and blue lines indicate morning and evening terminators, and the brown line indicates the subsolar meridian. Notice the absence of morning hazes and the presence of a fully visible band on the 22 July. (e) Areographic distribution of measured bands over a grey topographic map made from MOLA data (Smith et al., 1999) (different colors represent different observations). (f) Graph showing the Latitude - Local Time distribution of all the observed bands. Different band colors indicate different periods starting on 18 June (blue), 1 July (green), and 18 July (red). Grey areas represent the night.

## 4. Tracking Motions over the Southern Polar Region (SPR)

Previous wind measurements of the SPR in a period (Ls = 337°-10°) with no dust storms were obtained by Wang & Ingersoll (2003) who analyzed Mars Obiter Camera images finding velocities from 10 to 20 ms$^{-1}$. Here we use VMC images to track motions of the dust during the 2018 dust storm (Figure 3). Previous wind measurements associated to cloud features using VMC images can be found in Sánchez-Lavega et al. (2018b). We have measured 10 pairs of VMC images taken between 1 July and 3 August separated by 20-40 minutes at a spatial resolution of ~11km/px that allows us to retrieve velocities with an estimated error of 10 ms$^{-1}$. We use animations and blinking between the two images of a selected pair to identify moving features, and then point manually to their centers to track their motion. Sharpest aerosol features (suitable for cloud-tracking) are mostly found in regions at morning Local True Solar Time. We assume that features act as passive tracers, and that their motions mark the velocity of the underlying winds. Although other possibilities cannot be excluded, the short time interval in each pair makes advection the most likely cause of motion in features that do not change much in shape or area.

The main trend visible in our measurements is that features move towards the terminator (Figure 3). The highest velocities occur at the edge of the polar cap, where they reach up to 110 ms$^{-1}$ (Figure 3a and Supporting Information Fig. S5). At equal latitudes, velocities in the morning are higher than in the evening. Winds are slower over the polar cap, with typical velocities of 60 ms$^{-1}$ in the morning (reaching 80-100 ms$^{-1}$ only in exceptional cases), and still lower (~20-40 ms$^{-1}$) around the pole and in the afternoon side. It is quite remarkable that the pattern of the wind field changes with date. Figure 3a, corresponding to 1 July, shows no hint of circulation around the pole, and the wind vectors are not oriented in the direction of the bands. On the contrary, on the 18 July (Figure 3b) wind vectors are suggestive of a circulation around the pole (longitudes ~135°E and 225°E and latitudes ~70°S-90°S), although the lack of measurements in the evening side means we should exercise caution when making this inference. In addition, the vectors show some orientation along the arc band that extends from latitude 80°S and longitude 150° to latitude 60°S and longitude 250°. These contradictory measurements of motions between the two dates show the complexity of the polar dynamics under a global dust storm condition.

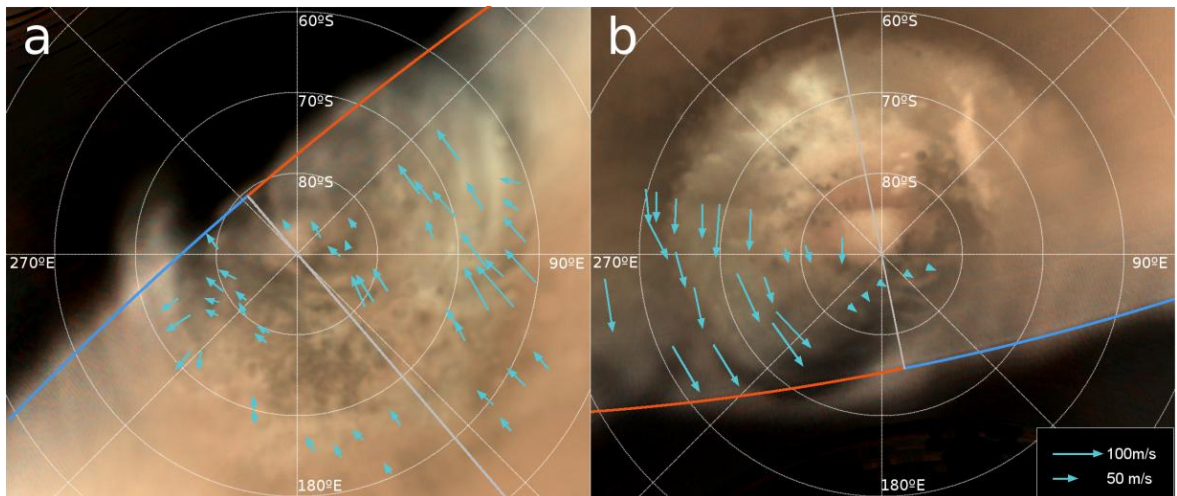

**Figure 3**. Velocity vectors retrieved tracking discrete features in two different dates: (a) 1 July, 17:20 U.T. and (b) 18 July, 14:40 U.T. Morning terminator, evening terminator, and subsolar meridian are indicated in the same format as in Figure 1.

## 5. Altitude Reached by Dust During GDS 2018

In addition to apocenter observations of the SPR, some VMC images were programmed to observe the limb of the planet from a closer distance. The orbit of MEx during this period allowed limb images of equatorial and northern latitudes with resolutions of ~ 7 km/px (Figure 4). At the limb, aerosols reflect sunlight against the dark background, and the maximum height and horizontal structure over the surface can be easily estimated once the geometry of the observation is determined (Sánchez-Lavega et al, 2018a). Figure 4 shows two examples of high altitude aerosols at the limb, most likely dust from the GDS according to the area occupied by the expanding storm (Sánchez-Lavega et al., 2019), reaching up to 60-70 km over the surface. Figure 4 also shows the location of all limb observations for the reported period and the dust top altitude. The maximum altitude, ~60 km, was observed on 19 June around Alba Patera. Scans of the limb show significant top altitude differences (from ~ 10 to 70 km) depending on the date and location, reflecting the complex dynamics involved in the GDS 2018.

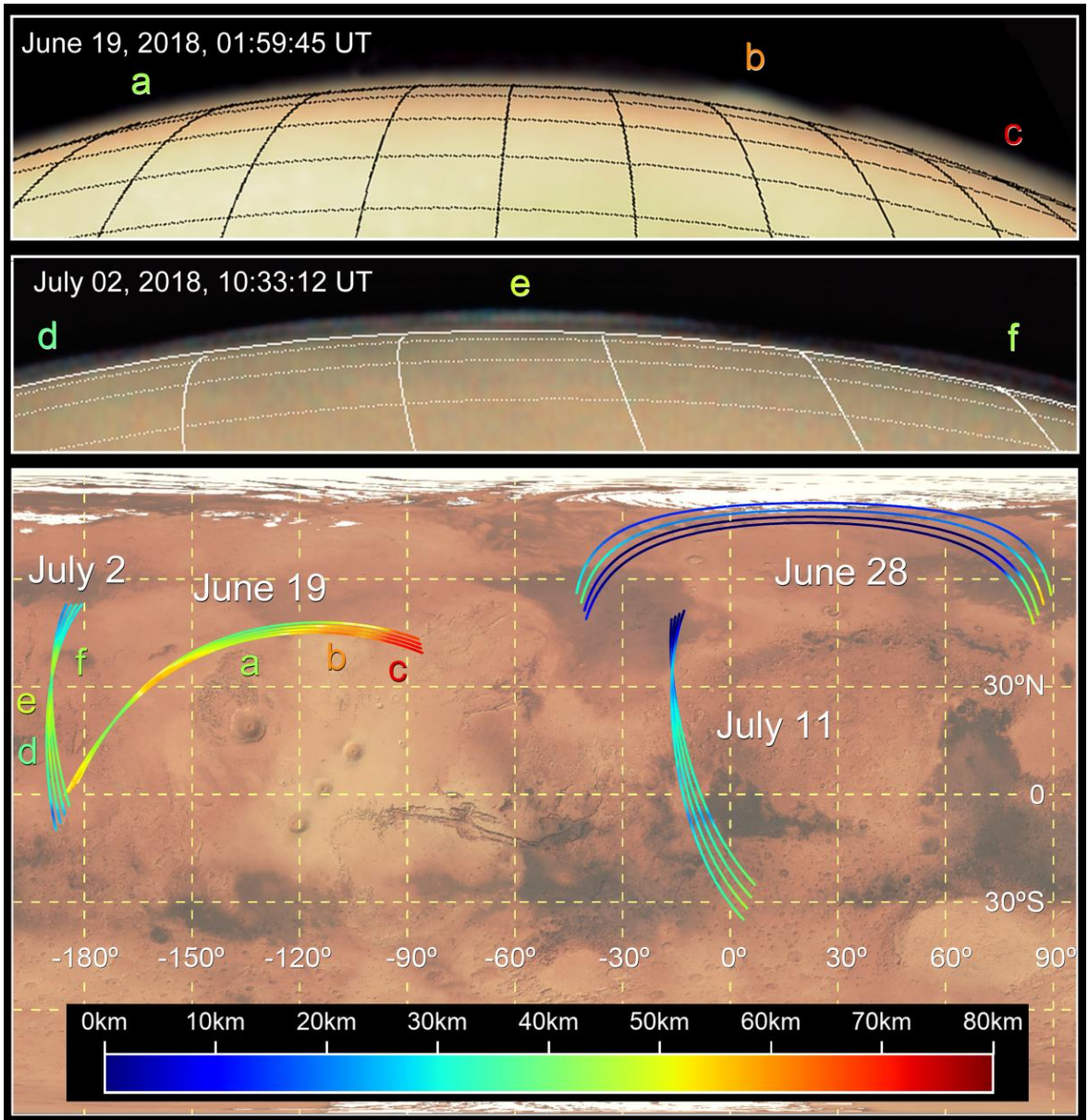

**Figure 4**. Upper two panels: Dust projected at limb in two different dates corresponding to a planet wide expansion of the GDS 2018 (Sánchez-Lavega et al., 2019). Lower panel: Location of the limb observations over a map of Mars. Top altitudes are indicated using a color code.

## 6. Comparison with Model Predictions

The Mars Climate Database (Forget et al., 1999; Millour et al., 2015) provides easy access to a statistical summary of results of the Global Climate Model of Mars developed by the Laboratoire de Météorologie Dynamique (LMD) under different dust and solar activity scenarios. In order to find out how the GDS affected the dynamics of the SPR, we have compared our measurements with the predictions from the model. Since the GDS developed under solar minimum, we considered two solar minimum scenarios: climatology (low dust) and dust storm. The dust storm scenario is not expected to describe the situation with accuracy, but it might be a first approach to the actual dynamics of the atmosphere during the dust storm.

For the orbital Solar Longitude and time of the day in Mars corresponding to the observation on 1 July at 17:20 U.T (Figure 3a) the MCD winds are shown in Figure 5. Since our cloud-tracking measurements lack altitude information, we present the velocity field from the model in the altitude range of 0-50 km, which is consistent with the altitude measurements in the previous section. Above 10km, the MCD predicts winds moving towards the terminator, mostly in the evening side; winds in the morning side are slower. These winds seem to result from a displacement of a predicted polar vortex in the direction of the morning side, relative to the terminator. In fact, a similar shift of the north polar vortex was observed in a reanalysis of a regional dust storm in MY 26, Ls~320º (Mitchell et al., 2015). The prediction of motions toward the terminator above 10 km is globally consistent with our measurements shown in Figure 3a, although our measured values do not agree with the predicted velocities, as MCD predicts faster winds in the evening side than in the morning side at any altitude over 10km, while we observe faster winds in the morning side.

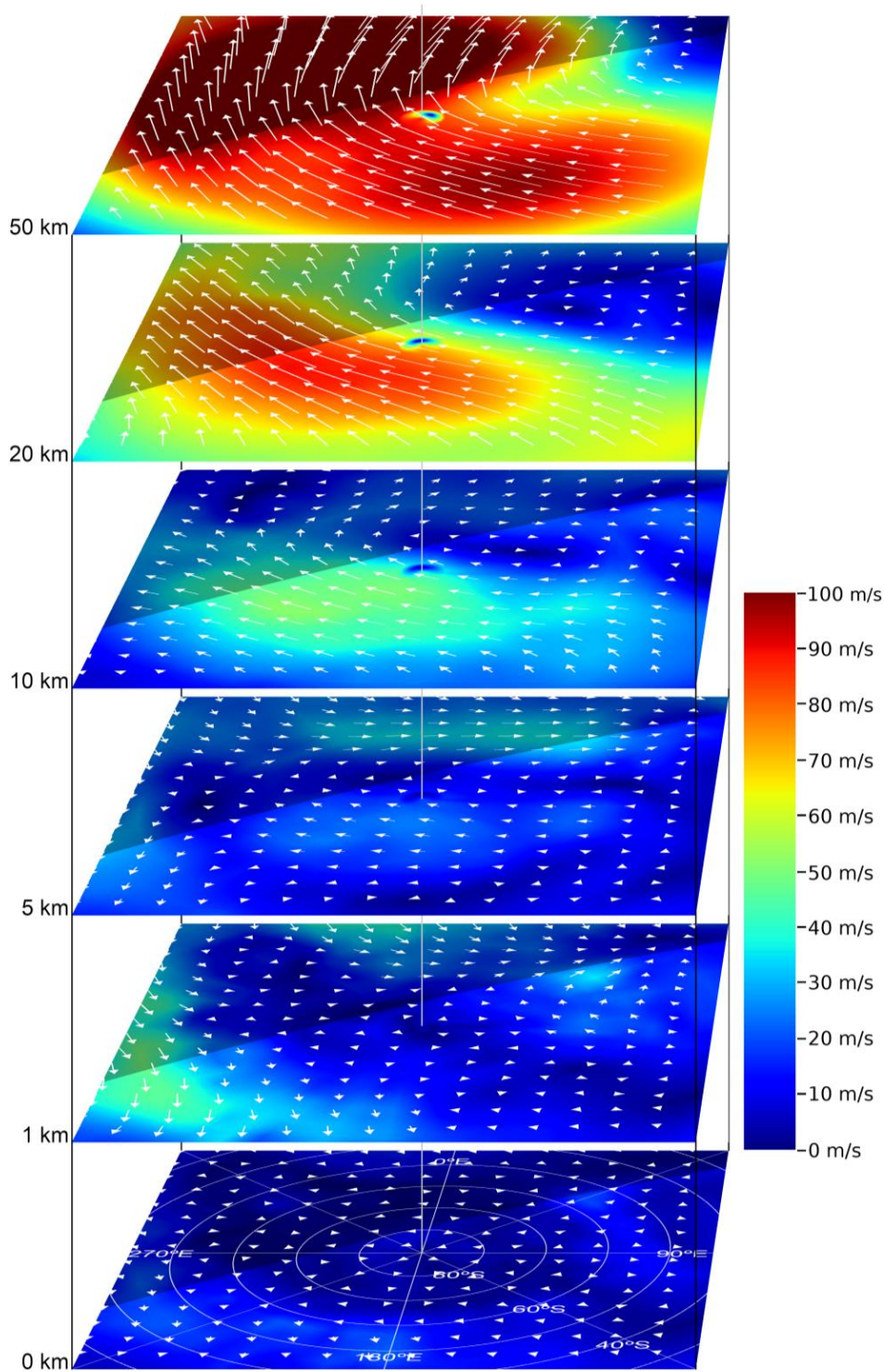

**Figure 5.** Wind field retrieved from MCD for dust storm solar minimum scenario, showing winds over the South Pole on 1 July at 17:10 (same time as for Figure 3a) at different altitudes. Equivalent figure for Figure 3b is available in Supporting Information. A white line in the maps shows the terminator. The darker area corresponds to the night side and the lighter are to day side.

## 7. Conclusions

The GDS 2018 penetrated the southern polar cap region, as already reported by Sánchez-Lavega et al. (2019), but dust coverage was not complete or homogeneous and we do not observe the south polar cap fully covered by dust in any image. In the north, the GDS expansion apparently stopped at ~ 45º-50ºN (Sánchez-Lavega et al., 2019). This asymmetric behavior of the GDS between the two poles can probably be explained by the presence of a strong polar vortex in the northern hemisphere, and a weak one in the south. This is in agreement with the observations that Martian polar vortices are stronger in the winter, and that the north polar vortex is stronger than the south polar one (Mitchell et al., 2015; Waugh et al., 2016).

The wind speeds and measured directions of the dust motions show differences when compared with the MCD predictions for a dusty scenario, particularly in the relative intensity of evening and morning winds, suggesting that the GDS 2018 altered the polar dynamics significantly. A proper account of the observed dust load and its spatial distribution in Martian Year 34 have been described elsewhere in this number (Montabone et al, 2019), and its inclusion in the LMD Mars GCM will probable lead to a better agreement. One possible interpretation of the circumpolar banding organization reported in section 3 is that it results from an alignment of dust masses by the winds blowing with the polar vortex. When we compare the bands with our measured winds (Figure 3a), or with the MCD wind field prediction at 10-50 km height (Supporting Information Figure S4), we do not find correspondence between the bands and the circulation, or even a clear presence of a polar vortex. As an alternative explanation, recent numerical simulations of the polar dynamics with a shallow water model, which include diabatic effects (Rostami et al., 2018), show that, under normal conditions, the potential vorticity field organizes in patches that sometimes form bands and arcs around the pole resulting from nonlinear instabilities in the flow.

A good understanding of polar dynamics during this major dust storm event, when the dust penetrates the polar area, requires additional observations to the ones presented here. Studies from other instruments will surely provide data on the dust mass load (Montabone et al, 2019), temperature, chemical composition and other physical properties that will serve as a reference input to numerical models of polar dynamics.

**Acknowledgments, Samples, and Data**

This work has been supported by the Spanish project AYA2015-65041-P (MINECO/FEDER, UE) and Grupos Gobierno Vasco IT-1366-19. JHB was supported by ESA Contract No. 4000118461/16/ES/JD, Scientific Support for Mars Express Visual Monitoring Camera. We acknowledge support from the Faculty of the European Space Astronomy Centre (ESAC)

VMC images are available at
https://archives.esac.esa.int/psa/#!Table%20View/VMC%20(Mars%20Express)=instrument